\documentstyle[preprint,aps,prb]{revtex}
\begin{document} 
\draft
\title {The Static, Dynamic and Electronic Properties of \\ Liquid Gallium
Studied by First-Principles Simulation}
\author{J.  M.  Holender and M.  J.  Gillan }
\address{Physics Department, Keele University \\ Keele, Staffordshire ST5 5BG,
U.K.}
\author{M. C. Payne}
\address{Cavendish Laboratory, Cambridge University\\Cambridge CB3 0HE, U.K.}
\author{A.  Simpson}
\address{Edinburgh Parallel Computing Centre, Edinburgh University\\
Edinburgh EH9 3JZ, U.K.}
\date{\today}
\maketitle
\begin{abstract} First-principles molecular
dynamics simulations having a duration of 8 ps have been used to study
the static, dynamic
and electronic properties of $\ell\mbox{\rm-Ga}$ at the temperatures  702\,K
 and 982\,K.
The simulations use
the density-functional pseudopotential method and the
system is maintained on the
Born-Oppenheimer surface by conjugate gradients relaxation.
The static structure factor and
radial distribution function of the simulated system agree very
closely with experimental data,
but the diffusion coefficient is noticeably lower than measured values.
The long simulations allow us to calculate the dynamical structure factor
$S(q,\omega)$.  A
sound-wave peak is clearly visible in $S(q,\omega)$ at small
wavevectors, and we present
results for the dispersion curve and hence the sound velocity, which is close
to the
experimental value.
The electronic density of states is very close to the free-electron form.
Values of the electrical conductivity calculated from Kubo-Greenwood formula
are in
satisfactory accord with measured data.
\end{abstract}
\pacs{ 61.20Ja, 61.25.Mv, 62.60.+v, 71.25.Lf}

\section{Introduction} Liquid gallium has been widely studied by a
 variety of experimental \cite{bel89,bro69,gin86,inu92,ber94}
  and theoretical methods
 \cite{haf84,haf90,tsa94,gon93}. The structure of the liquid is
 now well established over a wide range of temperatures but the
 knowledge of its dynamical properties, and particularly of its
 collective dynamics is in a much less satisfactory state.
 There is also a need for more detailed investigation of its
 electronic structure.  In the present work, we have undertaken
 first-principles  molecular dynamics (MD)
 simulations of $\ell$-Ga at two temperatures.
 These simulations are considerably longer than the only previous
 first-principles simulation \cite{gon93}, and this has allowed us to
 investigate the structure and electronic properties with greater
 statistical accuracy.  More importantly, it has allowed us to
 investigate the dynamical structure factor, which describes the
 dynamics of density fluctuations.

Collective fluctuations have previously been studied in a variety of
 metallic and insulating liquids.  Neutron-scattering experiments have
 shown that short wavelength sound-waves are readily observable in
 several liquid metals including Rb \cite{cop74}, Cs \cite{bod92} and
 Pb \cite{sod80}.  In general, oscillatory fluctuations can be
 observed for wavevectors up to over half the wavevector of
 the first peak in the static structure factor.  On
 the other hand, in some other liquids such as Ne \cite{bel75},
 sound waves are found to be strongly overdamped in this region of
 wavevectors.  Classical simulations of liquid rare gases
 \cite{lev73} and metals \cite{rah74} have revealed the same
 qualitative difference of behavior between the two types of liquid.  This
 difference has been traced to the fact that the short-range interatomic
 repulsion in simple metals is much softer than in rare gases \cite{han}.
There also seems to be a correlation with the ratio of specific heats
 $\gamma$, which is generally close to unity in liquid metals \cite{fab},
  but is
 considerably greater than unity in liquid rare gases \cite{fab} .  In this
 context, the behavior of $\ell$-Ga is puzzling.  Very recent inelastic
 neutron scattering measurements \cite{ber94} just above the melting
 point have failed to find oscillating density fluctuations, even
 though $\gamma$ for $\ell$-Ga has one of the lowest values among liquid
 metals.  Here, we report first-principles simulations only at high
 temperatures, so that a direct comparison with the recent inelastic
 data cannot yet be made, but we shall show that density oscillations
 are clearly visible in our system.  We shall suggest reasons for this
 apparent disagreement.

 Since the pioneering work of Car and Parrinello \cite{cp},
first-principles MD simulation has become
increasingly important in the study of liquids.  The basic idea of the
method is to calculate the total energy and the forces for any
arrangement of atoms by solving the equations of density functional
theory to determine the electronic ground state.  The great advantages
of this approach are that it completely avoids {\it ad hoc} assumptions
about the interactions between the atoms, and that it allows the
electronic properties of the liquid to be calculated within a unified
framework.  Simulations on $\ell$-Si \cite{sti89}, $\ell$-GaAs \cite{zha90},
$\ell$-Ga \cite{gon93}, $\ell$-CsPb \cite{wij94},
$\ell$-Ge \cite{kre94}, $\ell$-NaSn \cite{sch95} and a number of other
systems have been reported, and the structure of the simulated systems
generally agrees closely with experimental data.

One previous first-principle study of $\ell$-Ga has been reported
\cite{gon93}.  This was a rather short simulation at the single
temperature of 1000\,K, but was valuable in a number of ways.
It showed that the simulated system reproduces the known structure of
$\ell$-Ga
rather well.  It also allowed an investigation of the role of covalent
bonds in the liquid.  Lastly it gave useful new insights into the
electronic structure, and showed that the deep minimum in the density
of states at the Fermi level known to exist in crystalline $\alpha$-Ga
\cite{haf90}
disappears in the liquid.

The simulation technique we use differs in important ways from the one
 originally proposed by Car and Parrinello.  We do not treat the
 electronic degrees of freedom as fake dynamical variables, but
 instead we relax the electrons to the Born-Oppenheimer surface at
 each time-step by conjugate-gradients minimization \cite{gil89,pay92}.
   This has the
 advantages of allowing us to use much larger time-steps and also of
 avoiding the rather artificial use of thermostats needed for metals in
 the conventional Car-Parrinello technique.  We also use Fermi level
 smoothing, treating the occupation numbers as dynamical variables, as
 has been done by a number of other workers \cite{kre94,gil89,gru94}.

The main technical features of our simulations are explained in
Section \ref{tech}. Our results for the structural,
dynamical and electronic properties of $\ell$-Ga  are presented in
Section III. A discussion of our findings is given in Section IV,
and our conclusions are summarized in Section V.

\section{Techniques}
\label{tech}
\subsection{Method}We first summarize briefly the aspects of our simulation
technique that are standard \cite{df},
before describing in more detail the less
familiar features.
As usual in first-principles MD, only valence electrons are explicitly
represented, and it is  assumed that the core states are identical
to those in the free atom. In the present case, the Ga 4s and 4p
electrons are counted as valence electrons, and all more tightly bound
electrons are  counted as part of the core.
The interaction between valence electrons and the atomic cores is
represented by a norm-conserving non-local pseudopotential, which is
constructed
{\it ab initio via} calculations on the free atom (see below  for details).
The calculations are performed in periodic boundary
conditions, with the electronic orbitals expanded in plane waves.
In this expansion, all plane waves are included whose wavevector
$\bbox{G}$ satisfies $\hbar^2G^2/2m < E_{\rm cut}$ where
$E_{\rm cut}$ is referred  to as the plane-wave cutoff energy.
The calculations can (and in principle should) be taken to convergence
with respect to the size of the basis set by systematically increasing
$E_{\rm cut}$. The exchange-correlation term in the density
functional expression for  the total energy is represented by
the Local Density Approximation \cite{lda}. A simulation is performed
by making the ions
follow classical trajectories determined by the forces acting
on them, while the electronic subsystem remains in the ground state
at each instant (the Born-Oppenheimer principle). All these features
are entirely standard.

There are two well-known problems in the first-principles MD simulation
of metals. The first is that Kohn-Sham states can cross the Fermi level,
so that their occupation number passes discontinuously between zero and unity.
This implies that the wavefunctions of occupied states can change
discontinuously
and that the forces on the ions can do the same. Both kinds of
discontinuity  can
wreak havoc with the numerical implementation of the equations of motion.
The second problem is that the Fermi discontinuity leads to
the need for extensive (and expensive) sampling over the Brillouin zone.

It was recognized many years ago \cite{fu83} that both these problems can
be solved at the same time by smearing
out the Fermi discontinuity so that the occupation numbers pass
continuously from unity to zero over  a specified energy
interval.
It was also shown \cite{gil89,gil88,dev92} that this idea can be formulated
as a variational
principle in which  the quantity to be minimized has the form of a
thermodynamic free energy,
which depends both on the Kohn-Sham orbitals and their
occupation numbers.
This formulation is closely related to the Mermin density
functional theory for electron systems at finite temperature \cite{mer65}.
The idea of working with variable occupation numbers in a free-energy framework
has been expounded by a number of authors \cite{kre94,gru94,wen90},
and all that is needed
here is a note of some technical features peculiar to the present work.

In general, the free  energy functional can be represented as

\begin{equation}
\label{A}
A[{\{\psi_i\}},\{R_I\},\{f_i\}] =
E[\{\psi_i\},\{R_I\},\{f_i\}] - \alpha Q(\{f_i\})\ ,
\end{equation}
where $E$ is the
total-energy functional, which depends on the
Kohn-Sham orbitals $\psi_i$, their occupation numbers
$f_i$ and the ionic positions $R_I$;
the quantity $Q$ plays the role of an entropy, and  $\alpha$  specifies
the smearing  width.
In general, $Q$ can be taken to have the form:
\begin{equation}
Q(\{f_i\})\,=\,2\sum_i\zeta(f_i)\ .
\end{equation}
If $\alpha$ is chosen to be $k_BT$ and $\zeta(f)$ has the form:
\begin{equation}
\zeta(f)\,=\,-\,f\, \ln\,f\,-\,(1-f)\,\ln(1-f)\ ,
\end{equation}
then minimization with respect to the $f_i$ at constant
electron number yields the usual Fermi-Dirac distribution:
\begin{equation}
f_i\,=\,1/[\exp((\epsilon_i-\mu)/k_BT)+1]\ ,
\end{equation}
where $\epsilon_i$ are the Kohn-Sham eigenvalues and $\mu$ is
the chemical potential.
However, it has been pointed out that many other choices of $Q$
and hence many other
equilibrium distributions for $f_i$ are possible \cite{gil88,dev92}.
A disadvantage
of the Fermi-Dirac distribution is that for a given
energy width  the occupation numbers $f_i$
approach their  asymptotic values of 0 and 1 rather slowly.
Because of this, we prefer a distribution that approaches its asymptotic
values in a Gaussian manner rather than exponentially.
This is easily achieved by taking the dependence of $f$ on $\epsilon$
to be given by:
\begin{equation}
f(x)= \left\{ \begin{array}{ll} {1 \over
2}\sqrt{e}\,\exp[-(x+2^{-1/2})^2] & \mbox {if $x > 0$}\\ 1-
{1 \over 2}\sqrt{e}\, \exp[-(x-2^{-1/2})^2] & \mbox {if $x <
0$} \end{array} \right.
\end{equation}
where
\begin{equation}
\label{alpha}
x=(\epsilon-\mu)/\alpha\ .
\end{equation}
It is readily shown that this equilibrium distribution is
obtained if $\zeta$ is chosen to be:

\begin{equation}
\zeta (x) = {1 \over 2}\, \sqrt{e}\, |x|\, \exp[-(|x|+2^{-1/2})^2]+
      {1 \over 4} \sqrt{\pi e}\  {\rm erfc} (|x|+2^{-1/2})\ ,
\end{equation}
where
\begin{equation}
 x(f)= \left\{ \begin{array}{ll}
[\ln(e^{1/2}/2f)]^{1/2}-2^{-1/2} & \mbox {if $f < {1 \over 2}$} \\
2^{1/2}-[\ln(e^{1/2}/2(1-f))]^{1/2}  & \mbox {if $f > {1 \over 2}$}
  \end{array}
\right.\ .
\end{equation}

In the original Car-Parrinello method, the plane-wave coefficients
were treated as fake dynamical variables. For metals,
this method leads to serious problems because of the rapid transfer
of energy from the ions to the electronic degrees of freedom.
Because of this, we prefer to use the conjugate gradients approach
\cite{gil89,pay92},
in which the electronic subsystem is  brought to the ground
state  at every step. This  also has the advantage of allowing one to use
a much larger time step than in the standard method.

As has been stressed before \cite{gil89}, with fractional occupation numbers
the free  energy is minimized only when the orbitals are eigenstates of the
Kohn-Sham
Hamiltonian. This is to be contrasted with the situation for insulators,
where  it suffices that the orbitals span the occupied subspace.
Because of this, it is essential to perform explicit
subspace rotation so as to make the orbitals eigenstates.
The procedure we use for this
is essentially the same as that of Gillan \cite{gil89},
which is closely related to the methods described subsequently by
Grumbach {\it et al.} \cite{gru94} and Kresse and Hafner \cite{kre94}.

Briefly, the above features of our technique are implemented by the
following strategy which is based on
Ref.\ \onlinecite{gil89}.
The occupation numbers and wavefunction coefficients  are varied
together and we minimize all bands simultaneously.
We apply  the standard conjugate-gradients technique to the wavefunction
coefficients.
For occupation numbers we apply a simpler method. At a given iteration
we have occupation numbers $\{f_{i}\}$. These are used to calculate the
`electronic forces' and the KS Hamiltonian. From the diagonal
elements of the KS Hamiltonian we calculate new occupation
numbers $\{\tilde{f}_{i}\}$.
The changes in the occupation numbers are made according to:
\begin{equation}
f'_{i}\,=\,f_{i}+ \gamma\,(\tilde{f}_{i}-f_{i})\ .
\end {equation}
The free  energy A is bound to decrease for some  positive value
of $\gamma$.
This change is made simultaneously with the standard conjugate gradients
step and is followed by
 subspace rotation \cite{gil89}.
This cycle is repeated until the change in the free energy
during one cycle is smaller than some tolerance.

Special attention has to be  paid to
 bands
above the Fermi level, which have low occupation numbers.
These influence the total energy in two ways: directly and indirectly.
The direct influence comes from the explicit appearance of wavefunctions
of weakly occupied bands in the  total energy.
The  indirect influence arises from projection of forces and subspace
rotation.
Weakly occupied bands should not be allowed to vary
in an uncontrolled manner and it is highly desirable that
they should be close to the KS eigenstates.
There are always  a few bands with very small occupation number and their
direct
contribution to the  total free energy is  almost negligible.
Since our algorithm is based on free energy minimization,
normal conjugate gradients will have great difficulty in bringing
these bands close to KS eigenfunctions.
To achieve this, we must use  preconditioning.
We work with scaled wavefunction coefficients.
We find that scaling of all wavefunction coefficients by the factor
$f_i^{1/2}$ solves problems with weakly occupied bands.

The calculations have been done partly with the CASTEP code \cite{pay92}
on the Fujitsu VPX240 at Manchester, and partly with its
parallel version CETEP \cite{cetep} on the CRAY T3D at Edinburgh.
The codes have been extensively rewritten,
 partly to allow all-bands operation, in which
 all bands are are updated simultaneously during the conjugate-gradients
 search, and partly to introduce the variable occupation number
 technique described above.

\subsection{Computational details}

The norm-conserving pseudopotential for Ga was constructed
using the standard Kerker \cite{ker} method, the s and p components
being generated from the neutral $4s^24p^1$ configuration
and the d component from the ionized
$4s^{0.75}4d^{0.25}$ configuration. In the practical calculations,
the pseudopotential is represented in the Kleinman-Bylander
separable form \cite{kb} with the s-wave being treated as local,
and the
non-local parts of the pseudopotential being treated in real space
\cite{kin91}.
In the construction of  the pseudopotential and in the simulations,
 the exchange-correlation energy is represented
in the Ceperley-Alder form\cite{lda}.

We have tested the pseudopotential by calculations on the
$\alpha$-phase of crystalline Ga.
This is the stable crystal structure under ambient conditions, and has
a based-centered orthorhombic Bravais lattice with eight atoms in the
unit cell. In order to achieve high accuracy, we have used
the rather large plane-wave cutoff of 250\,eV and
a set of 32 k-points.
The calculated values of lattice parameters were 4.37\,\AA, 4.38\,\AA,
7.42\,\AA, and the
internal parameters were 0.07 and 0.16. The corresponding experimental values
are 4.511\,\AA, 4.517\,\AA, 7.645\,\AA, 0.078 and 0.1525
(Ref. \onlinecite{wyc64}).
The main deficiency of the calculations is clearly the error of ca. 3\%
in the lattice parameters. Essentially the same error was reported
by Gong {\it et al.} \cite{gon91} and we believe it arises from the LDA
approximation.
In all our calculations for the liquid we accept
that we are bound to make this error and
all distances are scaled accordingly when
comparison with experimental data is made.

Our calculations on $\ell$-Ga were  all performed on a system of 64
Ga atoms using a cubic repeating cell, with the density equal to the
experimental value at  each temperature  with the
above mentioned scaling of  distances.
The plane-wave cutoff for the liquid simulations
was taken to be 125\,eV, and $\Gamma$-point sampling
was used in the calculation of the (free) energy and the forces
at each time step.
The Verlet algorithm was used to integrate the equations of motion for
the atomic positions and velocities, with a time step $\Delta t$ of 3\,fs.
As explained above, the number of bands has to be taken greater than half
the number of valence  electrons, in order
to allow for  partial occupation.
We worked with 102 bands, which is six more
than would be needed for the 192 valence electrons if all bands
were fully occupied. The smearing width (the parameter
$\alpha$ in Eqn. (\ref{alpha})) was equal 0.2\,eV.

We initiated the system at 1000\,K.
 We equilibrated our system for about 10\,ps at
this temperature and then we collected data
over 8\,ps. The average temperature was 982\,K.
Then the system was slowly cooled at the rate 40\,K\,ps$^{-1}$
until the temperature of 700\,K was reached.
This was followed by a further  run of 8\,ps at 700\,K, the average
temperature in this run being 702\,K.

\section{Results} \subsection{Structural properties}
\label{stpr}
The structure of the simulated liquid can be compared directly with
 that of the real system through the static structure factor $S(q)$, which
 is measured in diffraction experiments.  This quantity is a measure
 of the intensity of density fluctuations as a function of wavevector
 {\bf q}, and is defined by:

\begin{equation}
\label{esf} S(q)=\bigl<|\hat \rho(\bbox{q})|^2 \bigr>\ .
\end{equation}
Here the dynamical variable $\hat \rho({\bbox{q}})$ representing the
Fourier component of the atomic density at wavevector $\bbox{q}$ is
given by:

\begin{equation}
\hat \rho(\bbox{q})=N^{-1 / 2}\,\sum_{i=1}^N \,\exp(i\bbox{q \cdot r}_i)\ ,
\end{equation}
where $\bbox{r}_i$ is the position of atom $i$ and $N$ is the number of
atoms in the system.  The angular brackets in Eqs. (\ref{esf})
denote the thermal average, which in practice is evaluated as the
time average over the duration of the simulation. In practical
calculations of $S(q)$, we also average over $\bbox{q}$ vectors having
the same magnitude.

Our results for $S(q)$ at 702\,K and 982\,K are compared in
Fig.~\ref{sf} with the neutron diffraction data of
Bellissent-Funel {\it et al.} \cite{bel89}.
We note that the latter data differ substantially from the
much older results cited in
the compilation by Waseda \cite{was}, which appear to be much less reliable.
The measurements in Ref.\ \onlinecite{bel89} were performed only
at 329\,K and
956\,K.  Our results at 982\,K are compared directly with the
experimental data at 956\,K, but to make comparison at 702\,K we have used
a linear interpolation of the experimental values at 329 and 956\,K.
The small size of our simulated system places a rather strong
limitation on the wavevector resolution of our results, and this is
partly responsible for their spiky form in the region of the first
peak.  This feature is also present in previous first-principles
simulations of liquids (see e.g., Refs\ \onlinecite{gon93,sti89,zha90}).
  Allowing for this, the agreement between the
simulated and experimental structure factors is close,
particularly away from the first peak.  The period, amplitude and
phase of the oscillations beyond ca.  4\,\AA$^{-1}$ are very well
reproduced.

It is also interesting to consider the low-$q$ limit of $S(q)$, which is
related to the bulk isothermal compressibility $\chi_T$ by the
relation
 \begin{equation}
 S(q \rightarrow 0)\,=\, nk_BT\chi_T\ ,
 \end{equation}
where $n$ is the number density.  From an extrapolation of our low-$q$
results at 702\,K, we obtain the estimate $S(q\rightarrow0)\,\approx\,0.012$,
which gives $\chi_T = 2.4\times10^{-11}\,$m$^2$N$^{-1}$.
This agrees fairly closely with the experimental value for
the adiabatic compressibility \cite{inu92}
 $\chi_S $which is $2.2 \times
10^{-11}\,$m$^2$N$^{-1}$. As $\chi_T/\chi_S = C_p/C_v = \gamma $ and
for $\ell\mbox{\rm-Ga}$ this value is \cite{fab} 1.1 the above comparison
of the isothermal and adiabatic compressibilities is well justified.
We shall return to the elastic
properties of the liquid when we discuss sound waves in section \ref{dynpr}.

The structure of the liquid can be seen more clearly from the radial
distribution function $g(r)$, and we compare simulated and experimental
\cite{bel89} results at 702\,K and 982\,K in Fig.  \ref{rdf}.  As before,
the~`experimental' curve is obtained by interpolation.
The close agreement between simulation and experiment in the region of
the first peak reflects the closeness of the structure factors beyond
ca.  4\,\AA$^{-1}$.  There are, however, slight discrepancies between
the $g(r)$ values beyond the first peak, and the reality of these
discrepancies is confirmed by the fact that they have the same form at
702\,K and 982\,K.  Since this region of $r$ is associated with the first
peak of the structure factor, where we have suggested an effect of system
size, it may well be that the small size of the system is responsible
for the discrepancies.

We have calculated the average coordination number defined in the
usual way as the average number of atoms within a distance $r_c$ of a
given atom, where $r_c$ is the distance at which $g(r)$ has its first
minimum.  Our calculated values at 702\,K and 982\,K are 8.7 and 9.1
respectively.  These values are essentially the same as the value of 9
that has been deduced from experimental measurements \cite{bel89}, as
would be expected from the close agreement between the simulated and
experimental $g(r)$. The coordination numbers in the crystalline
$\alpha$ and $\beta$ phases are 7 and 8 respectively, so that there is
a small but significant increase on going from solid to liquid.  This
is expected from the density increase on melting \cite{fab}.

\subsection{Dynamical properties}
\label{dynpr}
In order to study how the atoms diffuse in the liquid, we have
calculated the time-dependent mean square displacement (m.s.d) which we
denote by $\langle \Delta r(t)^2 \rangle$.  For large time interval $t$, the
asymptotic form of the m.s.d.  is expected to be:
\begin{equation}
\langle \Delta r(t)^2 \rangle \rightarrow B + 6\,D\,|t|\ ,
\end{equation}
where $B$ is a
constant and $D$ is the tracer diffusion coefficient.  In calculating
the m.s.d., we average in the usual way over all atoms in the system
and over time origins.  The interval between the origins is taken as
$\Delta t $, so that every step serves as an origin.

Our results for the m.s.d. at 702\,K and 982\,K displayed in
Fig.~\ref{diff} show that the atoms are diffusing rapidly, as expected.
For example, the plots show that at 982\,K the typical time taken for
an atom to travel the nearest neighbor distance of $2.7$\,\AA\ is
roughly 1.5\,ps.  As usually happens in highly mobile liquids, the
m.s.d.  rapidly reaches its asymptotic linear behavior, the transient time
being roughly 0.1\,ps.  From the asymptotic slope of the
m.s.d.  we obtain values for the diffusion coefficient of $3.3 \times
10^{-5}\,$cm$^2$\,s$^{-1}$
and $6.5 \times 10^{-5} $\,cm$^2$\,s$^{-1}$
at 702 and 982\,K respectively.  These are considerably
smaller than the rather old experimental values of
$9 \times 10^{-5} $cm$^2$s$^{-1}$
and $1.3 \times 10^{-4} $cm$^2$s$^{-1}$
at these temperatures \cite{bro69}.
We discuss later the significance of these discrepancies.

The m.s.d.  is a single-particle correlation function.  It is also of
considerable interest to examine the collective dynamics of density
fluctuations in the liquid, characterized by the intermediate
scattering function $I(q,t)$ and its Fourier transform the dynamical
structure factor $S(q,\omega)$.  The latter quantity is important
because it can be measured rather directly by inelastic neutron
scattering, and such measurements have recently been reported for $\ell$-Ga
\cite{ber94}.  The intermediate scattering function is defined as

\begin{equation}
 I(\bbox{q},t)= \Bigl<\hat \rho(\bbox{q},t) \cdot \hat \rho(\bbox{-q},0)\Bigr>\
,
\end{equation}
where $\hat \rho(\bbox{q})$ is the Fourier component of the density, as
before.  Note that $I(\bbox{q},t)$ is a real quantity which in an
isotropic liquid depends only on the {\it magnitude} of {\bf q}.

We have calculated $I(q,t)$ directly from its definition,
averaging both over time origins and over the orientation of {\bf q}.
The interval between time origins was taken to be $\Delta t$.
  Our results for a range of wavevectors at 702 and 982\,K are shown
  in Fig.~\ref{isf}.
The very similar form of the plots for the two temperatures confirms that
 the simulation runs are long enough to be statistically reliable.
 Note that at $t=0$,\ $I(q,t)$ becomes identical to the static structure
 factor $S(q)$, so that the systematic difference between the zero-time
 values at the two temperatures simply reflects the temperature
 dependence of $S(q)$, which we have already noted in Fig.~\ref{sf}.
The most significant feature of $I(q,t)$ is the pronounced oscillations
observed at low q, which rapidly become overdamped for $q> 1.8\,$\AA\,$^{-1}$.
These oscillations represent sound waves, as we shall see immediately.

The power spectrum of density fluctuations is described by the
dynamical structure factor defined by:

\begin{equation}
S(q,\omega)={1 \over \pi}\int^{\infty}_0\cos(\omega\,t)\, I(q,t)\,dt\ .
\end{equation}
We have performed the transformation using the  Welch window
function \cite{numrec}
 cutting off at maximum time 1\,ps.  Our results for
 $S(q,\omega)$ shown in Fig.~\ref{dsf} reveal two main features:  a
 peak centered at zero frequency representing decaying fluctuations
 and a finite-frequency peak representing oscillations.  These features
 have been observed many times before both in classical MD simulations
 of liquids and in inelastic scattering experiments. At small
 wavevectors,
 the central peak is associated with heat diffusion, but at the
 wavevectors we are dealing with here its interpretation becomes
   more complicated \cite{han}.
   As we have already seen from
 $I(q,t)$, the oscillatory fluctuations survive only for wavevectors up
 to 1.8\,\AA$^{-1}$, beyond which we are left only with the central peak.

Our results for $S(q,\omega)$ can be used to construct
a dispersion curve for sound waves
in the range of $q$ for which they can be observed. To do this, for each $q$
we have taken the frequency $\omega_{\rm max}$ at which the
sound-wave peak in
$S(q,\omega)$ has its maximum.
Fig.~\ref{disp} shows plots of $\omega_{\rm max}$ against $q$, on which we
have superimposed straight lines whose slope is equal to the experimental
adiabatic sound velocity in $\ell$-Ga \cite{inu92}. We note that
the shape of our dispersion curve is very
similar to experimental dispersion curves found for
Rb \cite{cop74}, Cs \cite{bod92} and
 Pb \cite{sod80}.

Note that a linear dispersion curve is expected only in the asymptotic region
of small $q$ where the wavelength is greater than all other
relevant lengths. The form of our dispersion curve indicates
that the lowest two of the wavevectors  available to us are within  this region
 to sufficient accuracy, so that we are justified in making a comparison
 with the measured sound velocity. It is not entirely clear whether
 the dispersion curve we observe should be associated
  with adiabatic or isothermal fluctuations.
  Strictly speaking, sound waves
  are adiabatic only at wavevectors for which the sound-wave
  frequency is much greater than the width of the Rayleigh peak, and this is
not
  obviously true in the present case.
  However, we have already pointed that
  the isothermal and adiabatic compressibilities should not differ more  than
  10\% in $\ell\mbox{\rm-Ga}$, and this suggests a  difference of the
  isothermal and adiabatic sound velocities of only 5\%.
  Our conclusion is that the good
  agreement of our dispersion curve with the measured sound velocity
  is genuine.
Close  agreement is, of course, not unexpected, since we saw in section
\ref{stpr} that the compressibility of the simulated system
 accords well with the known value.

\subsection{Electronic properties}

One of the important questions about $\ell$-Ga is the extent to which it
can be considered a free electron metal.  The most direct way of
studying this question is through the electronic density of states
(DOS), since deviations of this quantity from the free-electron
form are immediately apparent. It is well established that the DOS in
crystalline $\alpha$-Ga shows a deep minimum at the Fermi
energy \cite{haf90}, and it is of interest to know whether this feature
survives in the liquid.

It is important to recognize that adequate $k$-point sampling
 is  essential in calculating the DOS. In the first-principles MD simulations
themselves, $\Gamma$-point sampling is used in the calculation
of the total energy and the forces, but the direct use of
the Kohn-Sham eigenvalues
at the $\Gamma$-point would not be satisfactory for calculating the DOS.
Our procedure is to select a number of
configurations from the simulation run, and for each
one we have used the Kohn-Sham Hamiltonian generated
in the simulation to calculate the electronic eigenvalues at a set
of $k$-points. The DOS is than obtained by averaging over configurations
and $k$-points. The $k$-point
set we have used consists of the eight points ($\case1/8,\case{1}{8},
\case{1}{8}$), ($\case{1}{8},\case{1}{8},\case{3}{8}$), ($\case{1}{8},
\case{3}{8},\case{3}{8}$), ($\case{3}{8},\case{3}{8},\case{3}{8}$)
and cyclic permutations;
the
points are taken with equal weights.
Tests with other $k$-points indicate that this $k$-point set is perfectly
adequate. We have also tested explicitly the effect of
calculating the DOS with $\Gamma$-point sampling only, and
we find that this produces large spurious minima at
certain energies. We have found that good statistical
accuracy is already obtained  with a fairly small number of configurations,
and our results were obtained by averaging
over 5 configurations at each temperature.

We display in Fig.\ \ref{ds} our calculated electronic
DOS at 702 and 982\,K together with the free-electron
curve. It is clear that deviations from the
free-electron form are very small in both cases, and there is no
trace of the deep minimum at the Fermi level characteristic
of the $\alpha$-Ga structure.
This conclusion agrees with the findings of Hafner and Jank \cite{haf90}
which were based on
perturbation theory arguments, but does not agree well with the
first-principles
MD results of Gong {\it et al.} \cite{gon93},
which show
quite strong deviations from free-electron behavior.
We believe that this discrepancy is due
to the inadequate $k$-point
sampling used by Gong {\it et al.} as we discuss later.

We have calculated the frequency dependent electrical conductivity
 $\sigma(\omega)$
from the Kubo-Greenwood formula
\begin{equation}
\sigma(\omega)={{2 \pi e^2}
\over {3 m^2 \omega \Omega}} \sum^{occup}_i \sum^{empty}_j
\sum_{\alpha=x,y,z} |\langle\psi_i|\hat p_\alpha |\psi_j\rangle |^2
\delta(E_j-E_i-\hbar \omega) \ ,
\end{equation}
 where $\psi_i$ and $\psi_j $ ,
are the wavefunctions of states below and above the Fermi level
respectively and $E_i, E_j$ are the corresponding eigenvalues.
The operator $\hat p_\alpha$ represents the momentum of an electron
in Cartesian direction $\alpha$, and $\Omega$ is the volume of the
simulation cell.
This is an approximate formula, which treats the valence electrons
as propagating independently of one another.
As in the calculations of the DOS, Brillouin-zone
sampling is important,
and we have used the same $k$-point set as for the DOS.
Averaging is performed over 5 configurations at each
temperature.

We show our calculated $\sigma(\omega)$ in Fig.\ \ref{cond}.
Our main interest is in the d.c. conductivity and we have
included unoccupied states only up to 1\,eV above
the Fermi level. This means that $\sigma(\omega)$  is correctly
calculated only for  frequencies $\hbar\,\omega\,\le\,1\,$eV.
It should be noted that the statistical accuracy deteriorates
as $\omega$ goes to zero, but this does not prevent us from
making a reasonably reliable extrapolation to the d.c. value. Our
estimates for $\sigma(0)$
at 702\,K and 982\,K
are $2.5\times10^4\,\Omega^{-1}\,$cm$^{-1}$ and
$2.0\times10^4\,\Omega^{-1}\,$cm$^{-1}$.
The corresponding experimental values \cite{gin86} are
$3.0\times10^4\,\Omega^{-1}\,$cm$^{-1}$ and
$2.8\times10^4\,\Omega^{-1}\,$cm$^{-1}$ respectively.

\section{Discussion}

Our first-principles simulations of $\ell$-Ga are  considerably longer
than the only previous simulation reported so far \cite{gon93},
and we have been able to study its static, dynamic and
electronic properties in more detail than before. Our comparisons
of the static structure factor $S(q)$ and the radial distribution
function $g(r)$ at 702 and 982\,K with experimental data have confirmed
that the
structure of the simulated system agrees closely with that of the real
liquid. However, we have found noticeable discrepancies which appear as
spikes in $S(q)$ at certain wavevectors in the simulated system. This
effect is stronger at the lower temperature, and discrepancies
between the simulated and experimental $g(r)$ are also more significant
at this temperature, though it
should be remembered that the experimental $g(r)$ at 702\,K is actually
obtained by a rather large interpolation. We have suggested that
the spikes we find in $S(q)$ are associated
with the rather small size of our system, and we note that similar
effects have been seen in other first-principles simulations.
It would clearly be desirable to check this
point by repeating the simulations on larger systems, but we are not
in a position to do this at present.

Our results for the m.s.d.
$\langle \Delta r(t)^2 \rangle $ show typical
liquid-like behavior, with the asymptotic linear regime being
reached after only $\sim$ 0.1\,ps. The values of the diffusion
coefficient obtained from the asymptotic slope of the m.s.d. appear
to be somewhat low compared with the rather limited and old experimental data.
The highest temperature at which we can make a direct comparison is 702\,K,
where our simulated value appears to be too low by a factor of
$\sim$\,2.7. An extrapolation of the experimental values suggests
that at 980\,K
our simulated result is too low by a factor of $\sim$2.
There are two possible explanations for this. Either the experimental
data are unreliable, or  the simulations are suffering  from a systematic
error. If one wished to take the latter point  of view, one would note that
the spikes in $S(q)$ are indicative of  a spurious ordering, which
might have the effect of suppressing the atomic diffusion.
This again points to the desirability of studies on larger systems.
At the same time, we  believe there would be a case for repeating the
experimental measurements, before deciding
that the simulations are at  fault.

Our investigations of the collective dynamics of
$\ell$-Ga  have shown that
density
oscillations are clearly visible for  wavevectors q\,$\le$\,1.5\,\AA$^{-1}$,
both directly through the oscillations in $I(q,t)$ and through the
finite-frequency peaks in $S(q,\omega)$. We have shown that the
dispersion curves obtained from these peaks give  a  small-$q$
slope that agrees  very closely with the experimental sound velocity.
This observation is not at all surprising, given the well established
density oscillations in other liquid metals in the corresponding
wavevector range.
However, it raises important questions about  the behavior of
$\ell$-Ga, since Bermejo {\it et al.} \cite{ber94} failed to observe
density oscillations in $\ell$-Ga at just above the melting point
(actually at 330\,K) using inelastic neutron scattering.
We believe that the key to this apparent disagreement lies in the
large difference of temperatures employed
in the neutron-scattering measurements  and in our simulations.
It would be expected that the shear and longitudinal viscosities
would increase with decreasing temperature, so that sound waves
would be more heavily damped at lower temperatures.
Unfortunately, there seems to bo no experimental
information on the temperature dependence of the viscosities in
$\ell\mbox{\rm-Ga}$, but we  note that an increase of the viscosities
with decreasing temperature would be generally consistent with the
known decrease of diffusion coefficient at lower temperatures.
It is plausible that such an increase of viscosity
at low temperatures would lead to overdamping of sound waves
in the wavevector range observable by neutron scattering.
We frankly admit  that this explanation is speculative.
Resolution of this question clearly requires either extension of the
experiments to higher or extension of the simulations to lower
temperatures.

Our calculations of the DOS demonstrate that the electronic structure
is close to being free-electron-like at both temperatures studied.
In fact, our calculated DOS is much closer to the free
electron form
than the DOS reported by Gong {\it et al.} \cite{gon93}.
There is an important technical point here. As we have emphasized,
completely erroneous  results for the  DOS are obtained if adequate
sampling over the Brillouin zone is not performed, a point which has
been made before Kresse and Hafner \cite{kre94}.
But Gong {\it et al.} report that they have
calculated the DOS using $\Gamma$-point sampling only, and
we believe that this must cast serious doubt on their results. We note
that the close agreement between the electronic DOS and the
free electron form provides strong support for Hafner's perturbation
theory approach \cite{haf90}, in which the structure
of $\ell$-Ga and other metals has been treated by expanding the total energy
of the system to second order
in the pseudopotential starting from the free electron gas.
The satisfactory agreement of our calculated d.c. conductivity
with experimental values \--- the 20\% discrepancy that we find
is rather typical of what has been found in previous first-principles
work \--- provides useful confirmation that the electronic
structure
of our simulated system is close to that of the real liquid.

\section{Conclusions}

In conclusion, we have performed first-principles MD simulations
of $\ell$-Ga at two high temperatures.
The static structure  of the simulated system agrees closely with
that of the real liquid,
but there are slight discrepancies
which may arise from the small size  of the simulated system.
The diffusion coefficient  of the simulated system appears
to be too low  by somewhat over a factor of two.
Oscillating density fluctuations are clearly visible,
the associated sound  velocity being in excellent
accord with the known value. The failure to
observe sound waves in earlier  neutron-scattering experiments
at just above  the melting point may be  due  to the much higher viscosity
at low  temperatures. The electronic DOS is very close to the free
electron form,
and the calculated d.c. conductivity
is in satisfactory agreement with experiment.

\acknowledgements

The work of JMH is supported by EPSRC grant GR/H67935. The computations
were performed partly on the Fujitsu VPX240 at Manchester Computer
Centre under EPSRC
grant GR/J69974, and partly on the Cray T3D
at Edinburgh Parallel Computer Centre
using an allocation of time from the High Performance Computing Initiative
to the U.K. Car Parrinello consortium.
Analysis of the results was performed using distributed hardware
provided under EPSRC grants GR/H31783 and GR/J36266.
We are grateful to Dr. M. C. Bellissent-Funel for sending us
the numerical results of the neutron measurements of
the structure of liquid gallium.
Useful technical assistance from Dr. I. Bush at EPSRC Daresbury
Laboratory is gratefully acknowledged.

\begin{figure}
\caption{The static structure factor $S(q)$ of $\ell$-Ga at 702 and
982\,K.  Solid line and circles connected
by dotted line represent simulation and experimental
values \protect \cite{bel89} respectively.}
\label{sf}
\end{figure}

\begin{figure}
\caption{The radial distribution function $g(r)$ of $\ell$-Ga
 at 702 and 982\,K.
Solid line and circles represent simulation and experimental results
\protect \cite{bel89}  respectively.}
\label{rdf}
\end{figure}

\begin{figure}
\caption{Time dependent mean square displacement
$\langle \Delta r(t)^2 \rangle$ for $\ell$-Ga
calculated from simulations at 702\,K (solid line) and 982\,K (dotted line).}

\label{diff}
\end{figure}

\begin{figure}
\caption{The intermediate
scattering function $I(q,t)$ for $\ell$-Ga   calculated at
702\,K (solid line) and 982\,K (dotted line) for six different
wavevectors $q$.}

\label{isf}
\end{figure}

\begin{figure}
\caption{The dynamical structure factor $S(q,\omega)$ for
$\ell$-Ga calculated at
702\,K (solid line) and 982\,K (dotted line) for six different
wavevectors $q$.}

\label{dsf}
\end{figure}

\begin{figure}
\caption{The dispersion curve for sound waves in $\ell$-Ga at 702 and 982\,K.
Circles represent frequencies $\omega_{\rm max}$ at which $S(q,\omega)$
of the simulated system has its peak value. The straight line represents
the experimental
sound-wave velocity \protect \cite{inu92} at each temperature.}

\label{disp}
\end{figure}

\begin{figure}
\caption{Electronic density of states of $\ell$-Ga calculated
from simulations at 702\,K (solid line) and 982\,K (dotted line) compared
with the free-electron form (dashed line). The vertical line denotes
the Fermi energy.}

\label{ds}
\end{figure}

\begin{figure}
\caption{Frequency-dependent electrical conductivity $\sigma(\omega)$
 of $\ell$-Ga calculated from simulations at 702\,K
 (open diamonds connected by solid line) and 982\,K (filled diamonds
 connected by dotted line).
 Open and filled triangles
 show experimental values
 of d.c. conductivity \protect\cite{gin86} at the two temperatures.}

\label{cond}
\end{figure}

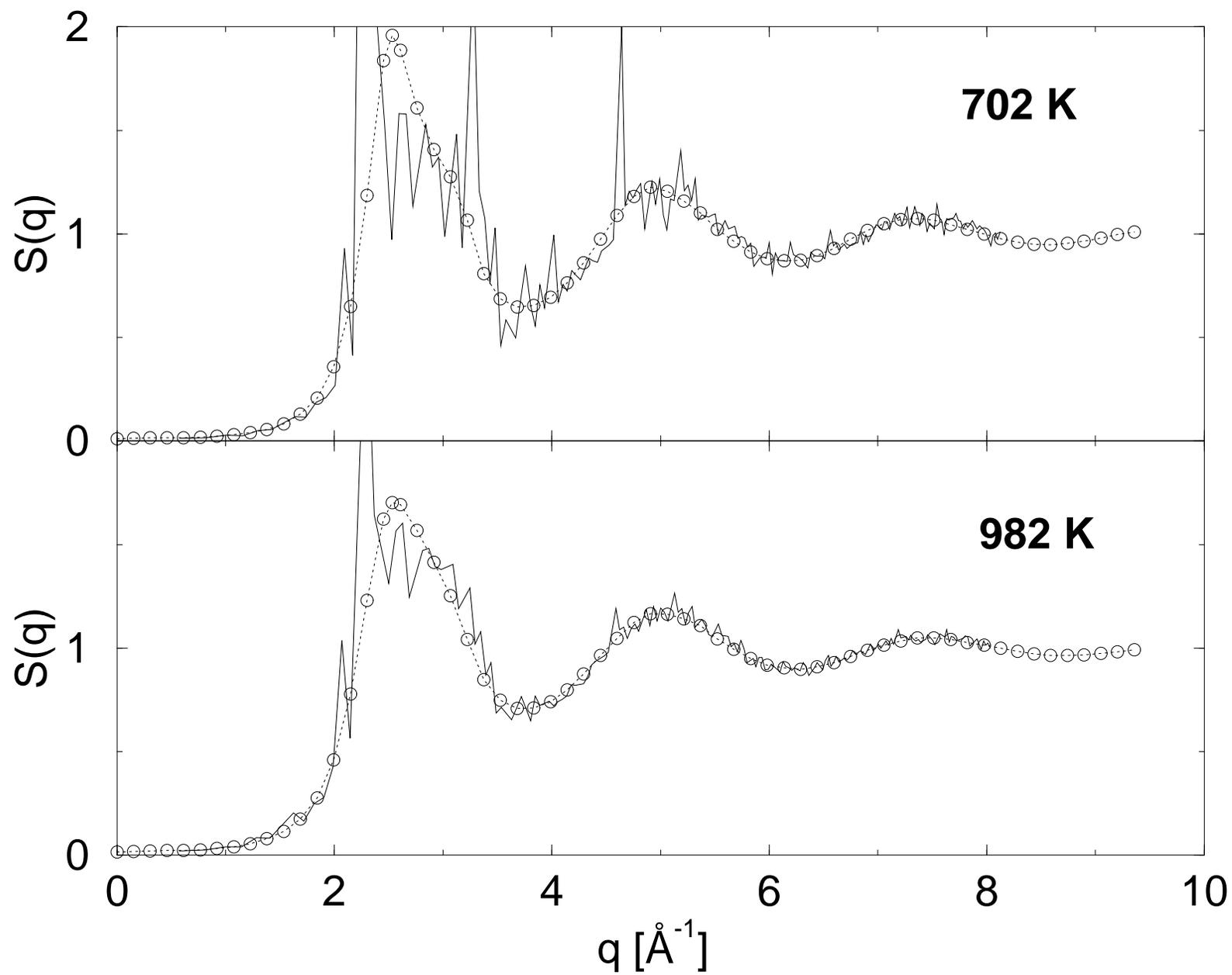
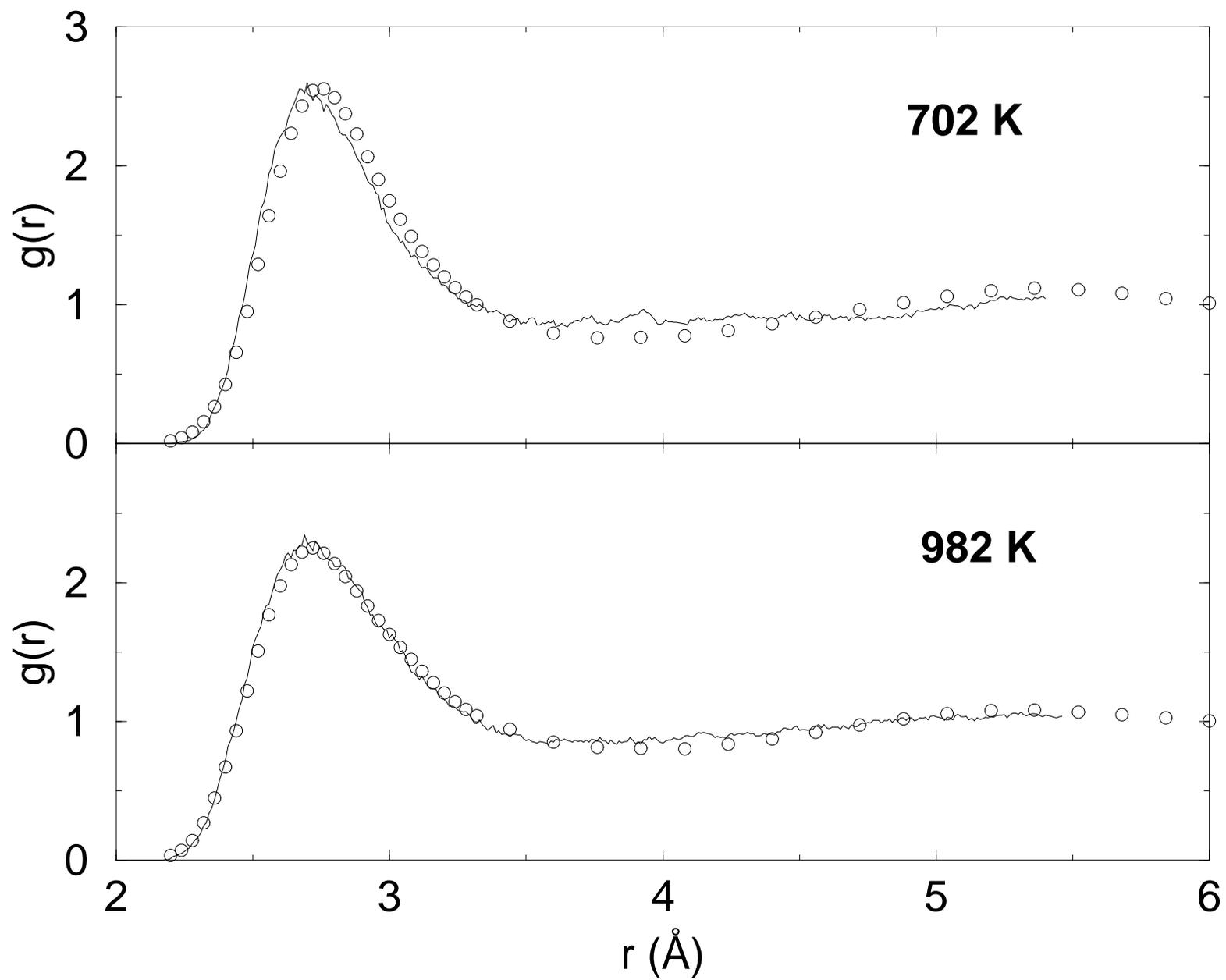
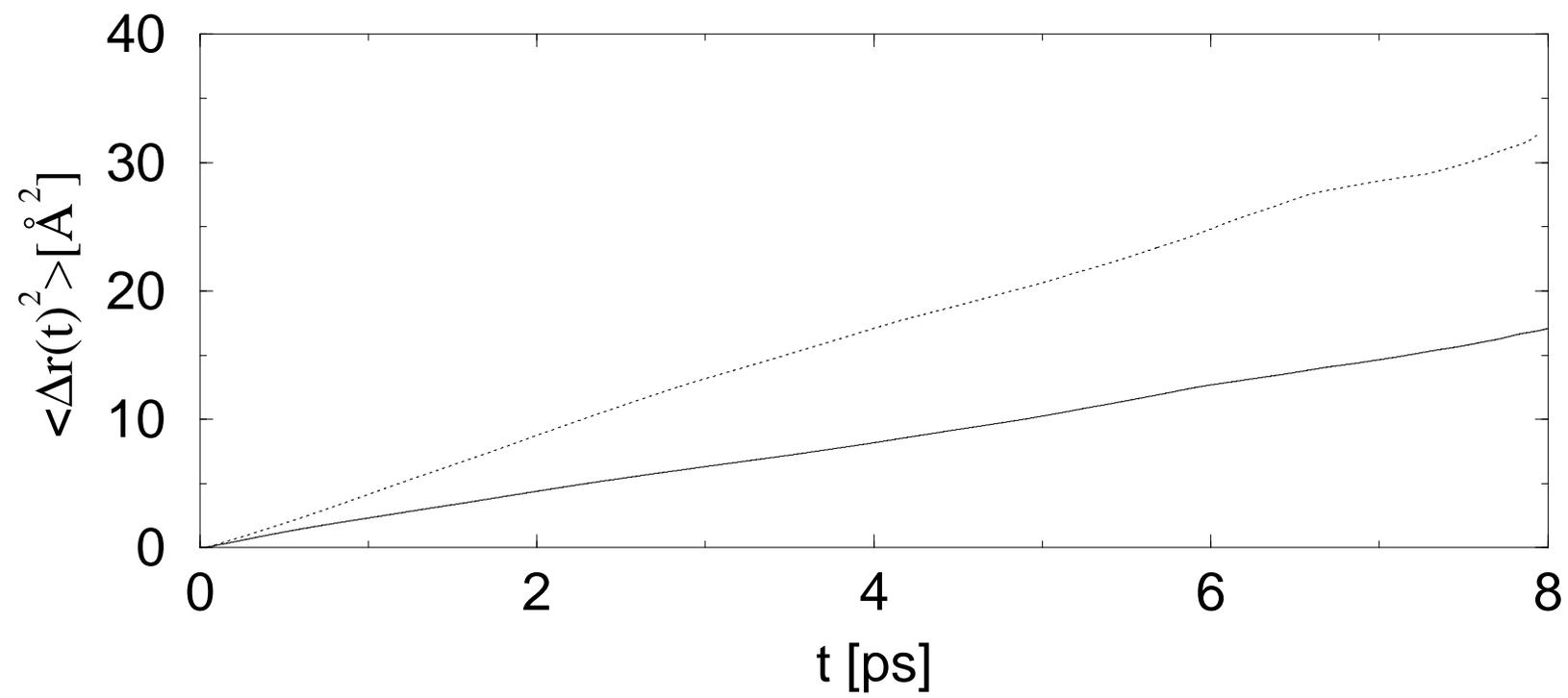
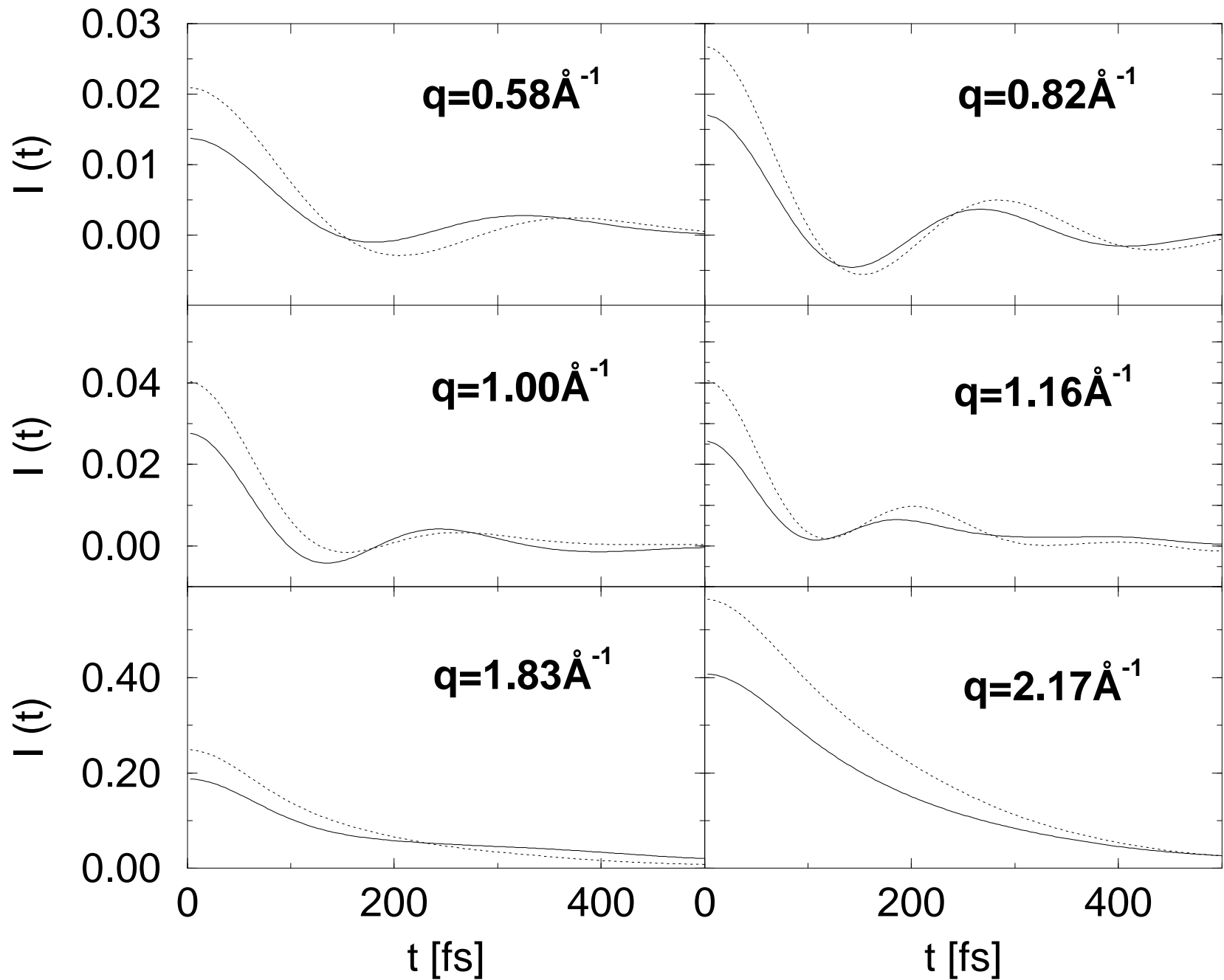
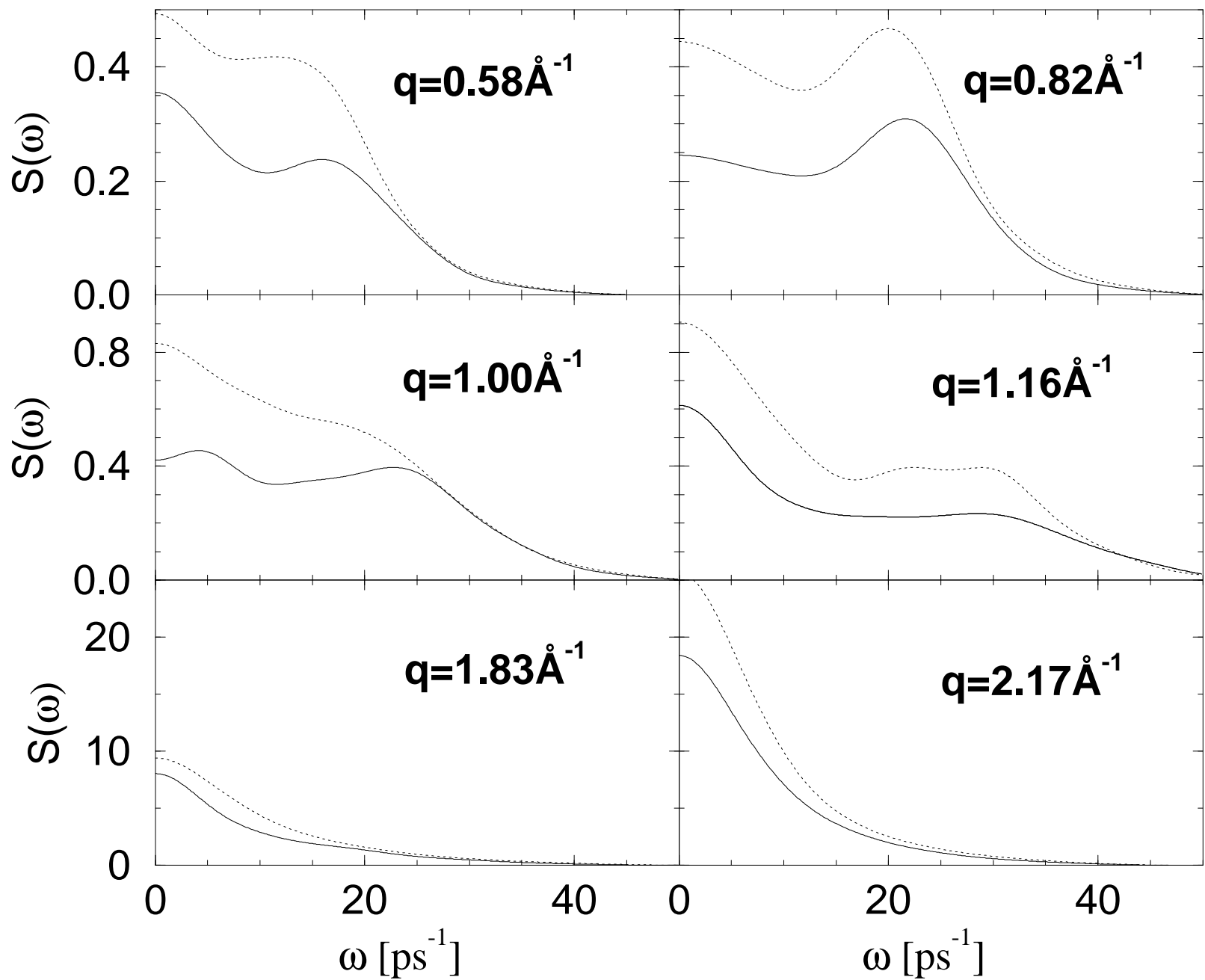
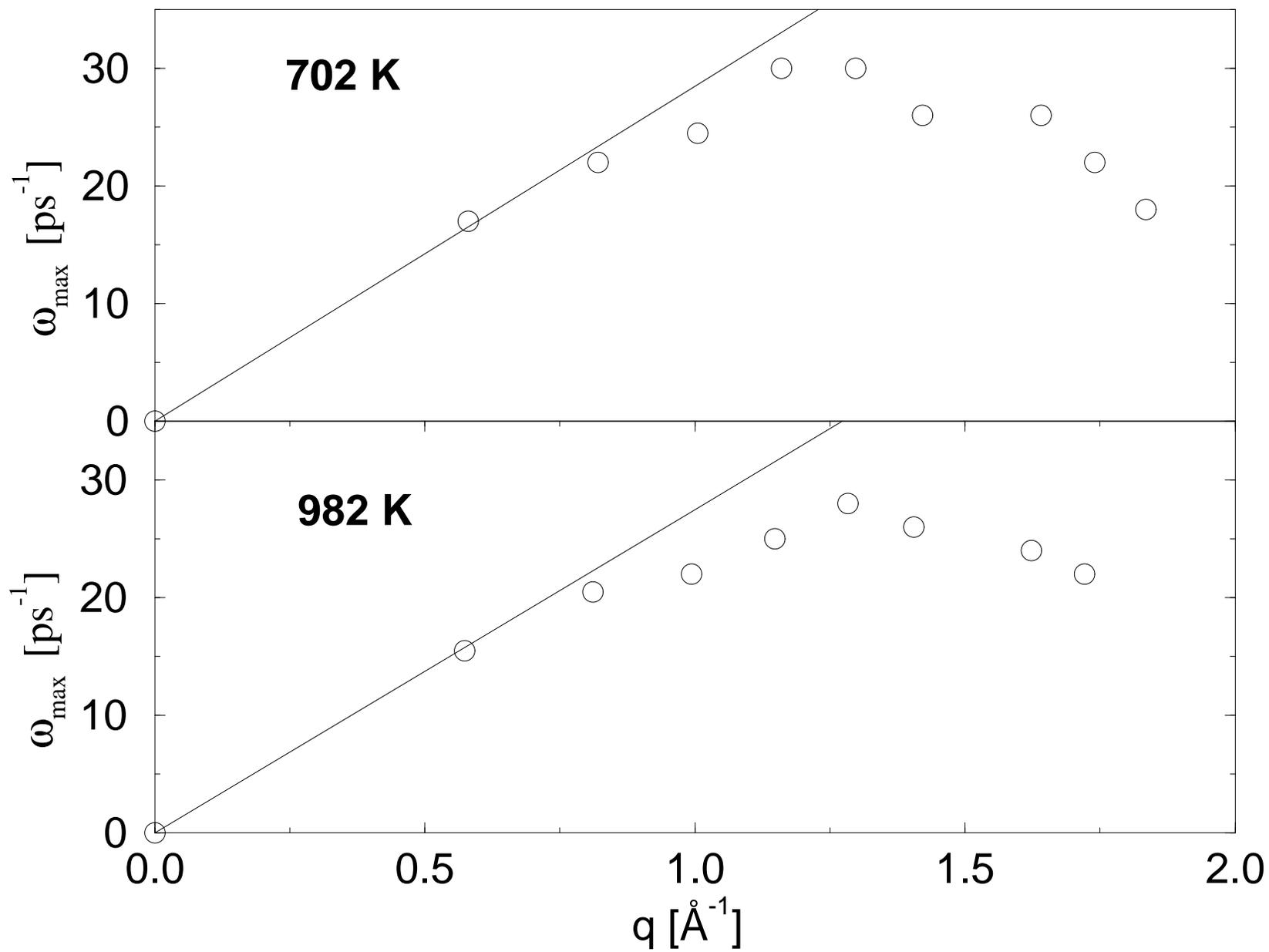
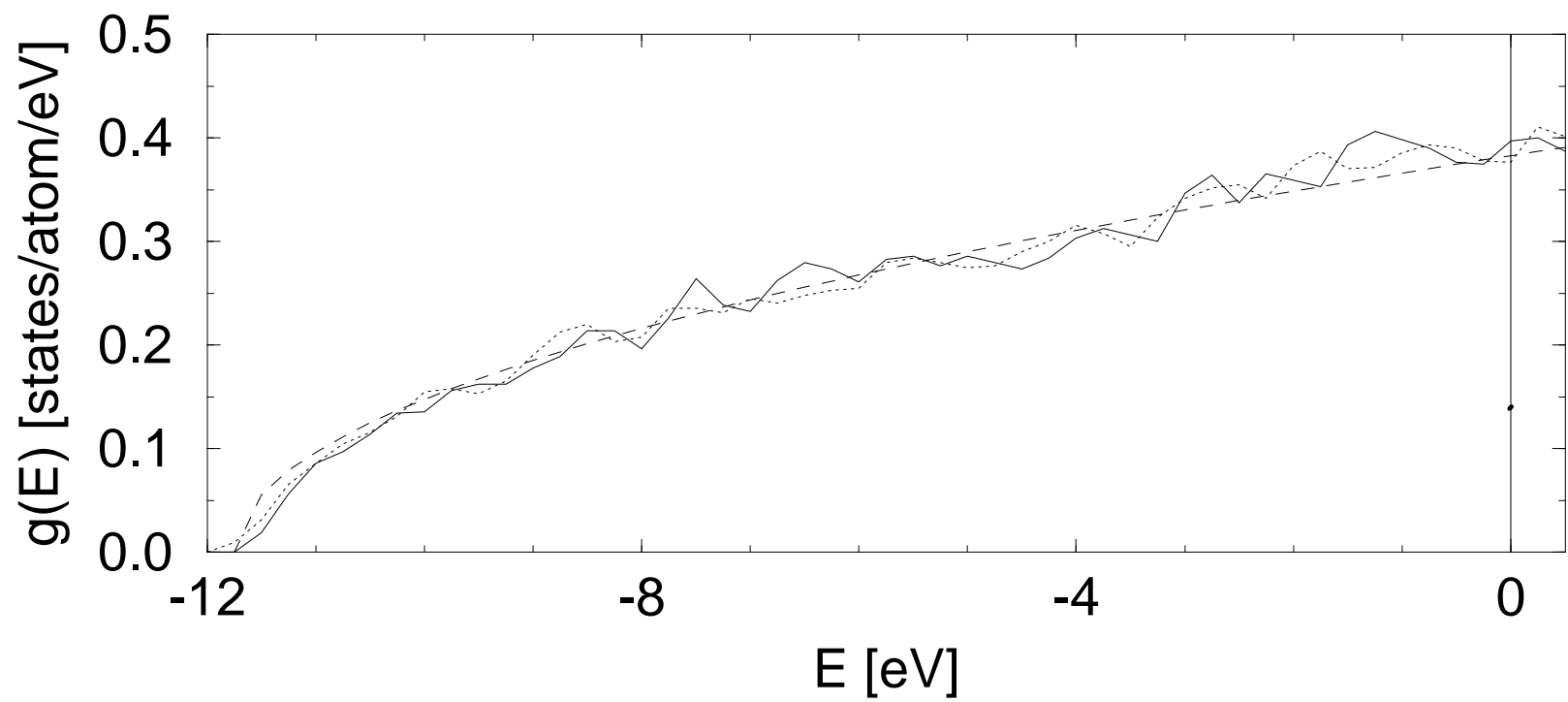
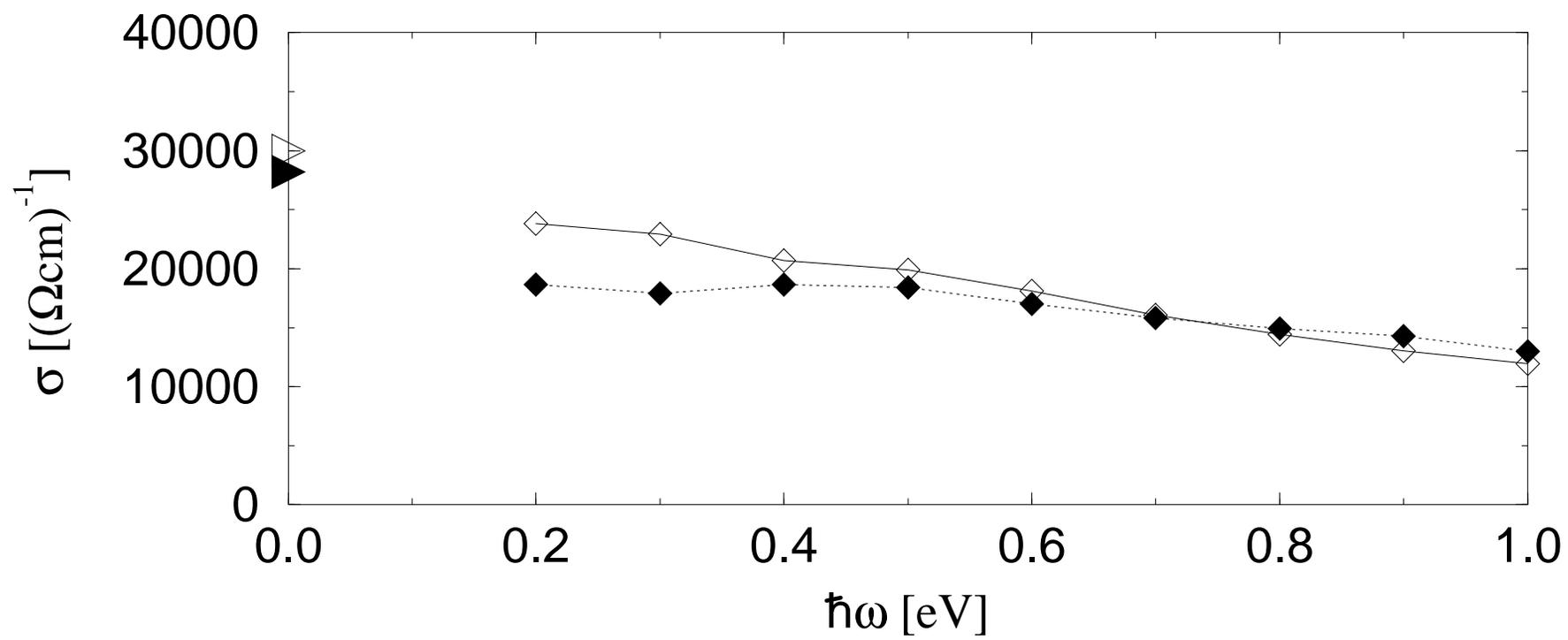
\end{document}